\documentclass[aip,graphicx,amsmath,amssymb,reprint]{revtex4-1}

\usepackage[version=3]{mhchem} 
\newcommand{\ssc}{\ensuremath{\mathrm{ss}}}

\newcommand{\app}{\ensuremath{\mathrm{app}}}

\usepackage{tikz} 
\usepackage{chemfig}
\usepackage{pgfplots}
\usepackage{tikzscale}
\usetikzlibrary{external}
\usetikzlibrary{shapes,arrows,positioning,patterns}
\usetikzlibrary{arrows.meta,bending}
\usepackage{changes} 

\renewcommand{\added}[1]{{\color{black}#1}}
\renewcommand{\deleted}[1]{}

\begin{document}

\title{Bruce-Vincent transference numbers from molecular dynamics
  simulations}

\author{Yunqi Shao} 
\author{Chao Zhang}
\email{chao.zhang@kemi.uu.se} 

\affiliation{Department of
  Chemistry-\AA{}ngstr\"om Laboratory, Uppsala University, Lägerhyddsvägen 1, P.
  O. Box 538, 75121 Uppsala, Sweden}

\date{\today}

\begin{abstract}
Transference number is a key design parameter for electrolyte materials used in electrochemical energy storage systems. However, the determination of the true transference number from experiments is rather demanding. On the other hand, the Bruce-Vincent method is widely used in the lab to measure transference numbers of polymer electrolytes approximately, which becomes exact in the limit of infinite dilution. Therefore, theoretical formulations to treat the Bruce-Vincent transference number and the true transference number on an equal footing are clearly needed. Here we show how the Bruce-Vincent transference number for concentrated electrolyte solutions can be derived in terms of the Onsager coefficients, without involving any extrathermodynamic assumptions. By demonstrating it for the case of PEO-LiTFSI system, this work opens the door to calibrating molecular dynamics (MD) simulations via reproducing the Bruce-Vincent transference number and using MD simulations as a predictive tool for determining the true transference number.

\end{abstract}

\maketitle


Transference number, defined as the fraction of current due to the migration of certain
ionic species, is an essential design parameter for the energy storage application of
electrolyte materials.  While important progresses have been made for the quest of the true transference number, notably with the combinations of concentration cell and steady-state measurements
\cite{2015_BalsaraNewman} or electrophoretic NMR
\cite{2019_RosenwinkelSchoenhoff}, its determination in polymer
electrolytes remains difficult in practice. 

The usage of steady-state currents for the determination of transport
coefficients dates back to experiments of Wagner on metal oxides and sulfides
\cite{1975_Wagner}. It is Bruce and Vincent\cite{1987_BruceVincent} who first derived
the equivalence of such measurements to determine the transference number of Li ions in polymer electrolytes. The method should give the true transference number in the limit of infinite dilution, however, it becomes approximated in concentrated solutions where the ion-ion correlations become non-negligible. Nevertheless, the Bruce-Vincent method \cite{1987_BruceVincent}
remains the most widely used one in the lab to gauge the transference number because of its simplicity. 

Computations of the true transference number $t^0_+$ from molecular dynamics (MD) simulations have come on the scene recently, which allow a direct comparison between theory and experiment~\cite{2022_ShaoGudlaEtAla,2022.Halat}.  However, due to the challenge to measure the true transference and the high accessibility of the Bruce-Vincent transference number $t_{+,\ssc}$ from experiment, it would be desirable to also obtain the Bruce-Vincent transference number directly from MD simulations.

In this Letter, we derive the $t_{+,\ssc}$ with the Onsager equations of ion transport and apply the method to the PEO-LiTFSI system using MD simulations. By clarifying the extrathermodynamic assumptions used previously, we show that the $t_{+,\ssc}$ has a well-defined connection with the Onsager coefficients and can be computed from MD simulations accordingly with a proper consideration of the reference frame (RF). Comparing the experiment and simulation for the PEO-LiTFSI system, we observe a consistent trend of its reduction with respect to its dilute limit, determined by the self-diffusion coefficients of ions.

\begin{figure}
    \centering
    \includegraphics[width=.95\linewidth]{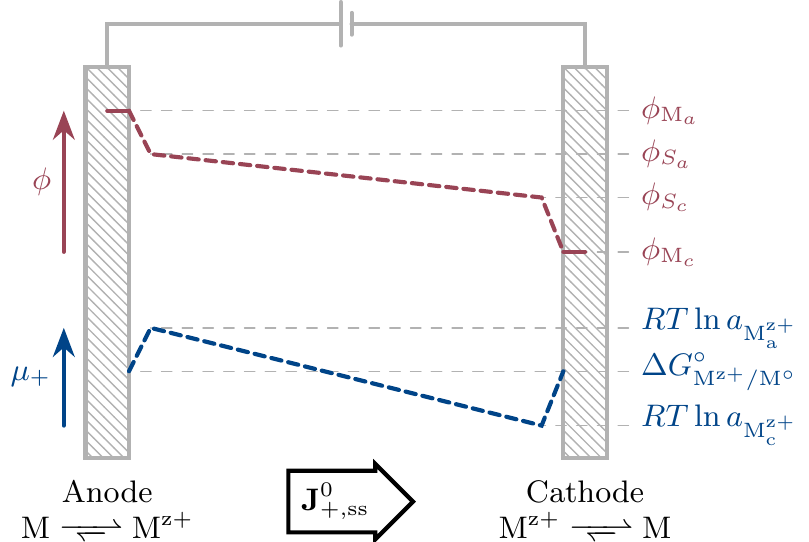}
    \caption{Illustration of the experiment setup; the dashed parts of potential
      $\phi$ and cation chemical potential $\mu_{+}$ cannot be measured, but their relation
      follows the Nernst equation. In this conceptualization, potential drops due to the interfacial/interphasial charge-transfer resistance at the steady-state \added{under the anion-blocking condition} are excluded.}%
    \label{fig:cell}
\end{figure}

The generic cell under consideration here is $\mathrm{M}_a \; | \; \mathrm{M}^{z+}\mathrm{X}^{z-} 
\;\mathrm{in \; polymer\; host}\; | \; \mathrm{M}_c$, as shown in Fig.~\ref{fig:cell}, where $a$ and $c$ denotes the anode and the anode respectively. Two electrodes are separated by a distance $d$ and the electrode reaction for cations is $
\mathrm{M}^{z+} + z e (\mathrm{M})\rightleftarrows \mathrm{M}$.

To begin the derivation of $t_{+,\ssc}$, we need to first introduce the driving forces, i.e. electrochemical potentials $\tilde{\mu}$ for both cation and anion at the steady-state:
\begin{align}
\tilde{\mu}_{+,\ssc} &= \mu_{+,\ssc} + zF \phi_\ssc \label{elechem_cat}\\
\tilde{\mu}_{-,\ssc} &= \mu_{-,\ssc} - zF \phi_\ssc \label{electrochem_ani}
\end{align}
, where $\mu_{+,\ssc}$ and $\mu_{-,\ssc}$ are the chemical potentials of cations and anions respectively, and $\phi_\ssc$ is the Galvani potential of the electrolyte solution. 

According to the Onsager theory of ion transport~\cite{onsager45}, the flux and driving forces are connected through Onsager coefficients and the reciprocal relation as follows:
\begin{equation}
    -\begin{bmatrix}
    \mathbf{J}^{0}_{+,\ssc} \\
    0
  \end{bmatrix}
  =
  \begin{bmatrix}
    \Omega^{0}_{++} & \Omega^{0}_{+-} \\
    \Omega^{0}_{+-} & \Omega^{0}_{--}
  \end{bmatrix}
  \begin{bmatrix}
    \nabla\tilde{\mu}_{+,\ssc} \\
    \nabla\tilde{\mu}_{-,\ssc}
  \end{bmatrix}
\label{OnsagerEq}
\end{equation}.

It is important to understand that the above relation is only valid in the solvent-fixed RF, which is the reason that the Onsager coefficients $\Omega^{0}_{+0}$ and  $\Omega^{0}_{-0}$ related to the
solvent species can be omitted.  The flux of anions is set to be zero in Eq.~\ref{OnsagerEq} by applying the \added{anion-blocking condition at the} steady-state\cite{1987_BruceVincent, 1988.Watanabe}, this allows us to express the flux of cations just using the electrochemical potential of cations alone:
\begin{equation}
  \mathbf{J}^{0}_{+,\ssc}
  = -\left(\Omega^{0}_{++}-\left(\Omega^{0}_{+-}\right)^{2}/\Omega^{0}_{--}\right)%
    \nabla\tilde{\mu}_{+,\ssc} 
\end{equation}.

The initial current density due to the migration of both cation and anions is \begin{align}
\label{external_potential}
    \mathbf{i}_\mathrm{tot} & = -\sigma \nabla V \\
     & = -z^2F^2(\Omega^{0}_{++}+\Omega^{0}_{--}-2\Omega^{0}_{+-})\nabla V 
\end{align}
, where $\Delta V = -d\nabla V $ is the applied potential and $\sigma$ is the ionic conductivity. It is clear from Eq.~\ref{external_potential} that the applied potential $\Delta V$ in this context stands for the one after excluding any potential drop due to the charge-transfer resistance at the interface. A similar procedure is also used in experiment by subtracting the $iR_\mathrm{ct}$ term~\cite{Evans:1987ho, 1988.Watanabe}.  

Then, the\deleted{n} Bruce-Vincent transference $t_{+,\ssc}$ as the ratio between the steady-state current (density) $\mathbf{i}_{+,\ssc}$ and the initial current (density) $\mathbf{i}_\mathrm{tot}$ can be expressed as follows:
\begin{align}
    t_{+,\ssc}  &= \frac{\mathbf{i}_{+,\ssc}}{\mathbf{i}_\mathrm{tot}}  = \frac{zF\mathbf{J}^{0}_{+,\ssc}}{\mathbf{i}_\mathrm{tot}} \\
    & = \left(\frac{\Omega^{0}_{++}-\left(\Omega^{0}_{+-}\right)^{2}/\Omega^{0}_{--}}{\Omega^{0}_{++}+\Omega^{0}_{--}-2\Omega^{0}_{+-}}\right) \frac{\nabla\tilde{\mu}_{+,\ssc} }{\nabla (zFV)}  \label{tss_eq1}
\end{align}.

Therefore, the key step for obtaining $t_{+,\ssc}$ is to establish the relationship between the change in the electrochemical potential of cations at the steady-state $\Delta \tilde{\mu}_{+,\ssc}$ and the applied potential $\Delta V$.

Following the notations from Fawcett~\cite{2004_Fawcett}, the Galvani potential difference at the $\mathrm{M}_a \; |$  solution interface $_{S_a}\Delta_{\mathrm{M}_a}\phi$ is defined as:
\begin{align}
    _{S_a}\Delta_{\mathrm{M}_a} \phi &=  \phi_{\mathrm{M}_a} -\phi_{S_a}\nonumber\\
    &= -\frac{\Delta G^{\circ}_{\mathrm{M}^{z+}/\mathrm{M}^\circ}}{zF} + \frac{RT}{zF}\ln a_{\mathrm{M}_a^{z+}}
\end{align}
, where $\Delta G^{\circ}_{\mathrm{M}^{z+}/\mathrm{M}^\circ}$ is the standard free energy for the reduction reaction of $\mathrm{M}^{z+}$, and $a_{\mathrm{M}_a^{z+}}$ is the activity of $\mathrm{M}^{z+}$ near the anode.  

The same applies to the cathode side with the Galvani potential difference $_{S_c}\Delta_{\mathrm{M}_c}\phi$ as
\begin{align}
    _{S_c}\Delta_{\mathrm{M}_c} \phi &=  \phi_{\mathrm{M}_c} -\phi_{S_c} \nonumber \\
    &= -\frac{\Delta G^{\circ}_{\mathrm{M}^{z+}/\mathrm{M}^\circ}}{zF} + \frac{RT}{zF}\ln a_{\mathrm{M}_c^{z+}}
\end{align}.

Then, the potential difference between two electrodes $_{\mathrm{M}_c}\Delta_{\mathrm{M}_a}\phi$ can be obtained by combining the two equations above:
\begin{align}
_{\mathrm{M}_c}\Delta_{\mathrm{M}_a}\phi & =\,_{\mathrm{M}_c}\Delta_{S_c}\phi\, + \,_{S_c}\Delta_{S_a}\phi + \,_{S_a}\Delta_{\mathrm{M}_a}\phi   \nonumber\\
& = \frac{RT}{zF}\ln \left(a_{\mathrm{M}_a^{z+}}/a_{\mathrm{M}_c^{z+}}\right) + (\phi_{S_a} - \phi_{S_c})
\label{eq:nernst}
\end{align}. It is worth noting that $_{\mathrm{M}_c}\Delta_{\mathrm{M}_a}\phi$ does not account for any potential drop developed at the electrode-electrolyte interface/interphase due to the charge-transfer resistance, which allows us to apply the Nernst equation to the Galvani potential difference here. 

By identifying $_{\mathrm{M}_c}\Delta_{\mathrm{M}_a}\phi = \Delta V$,  $\Delta \mu_{+, \ssc} = RT\ln (a_{\mathrm{M}_a^{z+}}/a_{\mathrm{M}_c^{z+}})$, and $\Delta \phi_\ssc = \phi_{S_a} - \phi_{S_c}$ and applying the definition Eq.~\ref{elechem_cat} of the electrochemical potential of cations,  the above equation can be rewritten as follows:
\begin{align}
    zF\Delta V &= \Delta \mu_{+, \ssc} + zF\Delta\phi_\ssc\\
    & = \Delta \tilde{\mu}_{+, \ssc}\label{V_mu_equality}
\end{align}.

Plugging Eq.~\ref{V_mu_equality} into Eq.~\ref{tss_eq1}, we get the following expression
\begin{equation}
    t_{+,\ssc}  = \frac{\Omega^{0}_{++}-\left(\Omega^{0}_{+-}\right)^{2}/\Omega^{0}_{--}}{\Omega^{0}_{++}+\Omega^{0}_{--}-2\Omega^{0}_{+-}}
    \label{tss_eq2}
\end{equation},
which is the main result of this work. 

For the infinitely dilute solution, $\Omega^{0}_{+-}$ becomes zero and there are also no correlations among the same type of ions. Therefore, one recovers the apparent transference number $t^s_\app$ which only depends on the self-diffusion coefficients $D^\mathrm{s}$, i.e., 
\begin{equation}
    \lim_{r\rightarrow 0}  t_{+,\ssc} = t_{+,\app} = \frac{D_+^\mathrm{s}}{D_+^\mathrm{s} + D_-^\mathrm{s}}
\end{equation}.

Before showing how to compute the Onsager coefficients from MD simulations and taking care of their RF-dependence, it is necessary to make a connection of Eq.~\ref{tss_eq2} with previous works~\cite{2015_BalsaraNewman, 2016_WohdeBalabajewEtAl}, where similar results were either implied or indicated but with seemingly rather different assumptions.  

The point for discussion is whether the assumption regarding the relationship between the chemical potentials of cations and anions matters or not.  For this purpose, consider the general case where $\mu_{+} = x\mu_{-}$. The linear relation between the steady-state flux
and the driving forces \added{under the anion-blocking condition} reads:
\begin{equation}
    - \begin{bmatrix}
        \mathbf{J}_{+,\ssc} \\ 0 \\ 0
    \end{bmatrix}   
     = \mathbf{M}
    \begin{bmatrix}
        \nabla\phi_{\ssc}\\
        \nabla\mu_{+\added{,\ssc}} \\
        \nabla\mu_{-\added{,\ssc}}
    \end{bmatrix} 
\end{equation}
with:
\begin{equation}
    \mathbf{M} =\begin{bmatrix}
        \Omega^{0}_{++} & \Omega^{0}_{+-} & 0 \\
        \Omega^{0}_{+-} & \Omega^{0}_{--} & 0 \\
        0 & 0 & 1
    \end{bmatrix}
    \begin{bmatrix}
        zF & 1 & 0 \\
        -zF & 0 & 1 \\
        0 & 1 & -x
    \end{bmatrix}
\end{equation}.
 
 Given that the only
non-zero term on the left-hand side is $\mathbf{J}_{+,\ssc}$, driving forces can be
represented in terms of $\mathbf{J}_{+,\ssc}$, $x$ and $\Omega$ by inverting the
linear relation, which leads to:
\begin{align}
  \begin{split}
    \nabla\phi_{\mathrm{ss}}
    &= -\frac{\mathbf{J}^{0}_{+,\ssc}}{zF}%
      \frac{x\Omega^{0}_{+-}+\Omega^{0}_{--}}%
{(x+1)\left(\Omega^{0}_{++}\Omega^{0}_{--}-\left(\Omega^{0}_{+-}\right)^2\right)} \\
    \nabla\mu_{-\added{,\ssc}}
    &= -zF\frac{\Omega^{0}_{+-}-\Omega^{0}_{--}}%
      {x\Omega^{0}_{+-}+\Omega^{0}_{--}}\nabla\phi_{\mathrm{ss}}\\
    \nabla\mu_{+\added{,\ssc}}
    &= x \nabla\mu_{-\added{,\ssc}}
  \end{split}%
  \label{eq:forces}
\end{align}.

Eq.~\ref{eq:forces} seems complicated but one can verify that they reduce
to the same set of equations reported in the literature under a specific choice of $x$, namely, $x=0$ as in
Ref.~\citenum{2015_BalsaraNewman} and $x=1$ as in
Ref.~\citenum{2016_WohdeBalabajewEtAl}. 

Importantly, the following two combinations of
the driving forces are $x$-independent:
\begin{align}
 \label{eq:mig}
  \nabla\mu_{+\added{,\ssc}}+zF\nabla\phi_{\mathrm{ss}}
  &= - \frac{\Omega^0_{--}}{\Omega^{0}_{++}\Omega^{0}_{--}-\left(\Omega^{0}_{+-}\right)^2} \mathbf{J}^{0}_{+,\ssc} \\
  \label{eq:diff}
  \nabla\mu_{+\added{,\ssc}} + \nabla\mu_{-\added{,\ssc}}
  &= \frac{\Omega^{0}_{+-}-\Omega^{0}_{--}}%
    {\Omega^{0}_{++}\Omega^{0}_{--}-\left(\Omega^{0}_{+-}\right)^2} \mathbf{J}^{0}_{+,\ssc}
\end{align}.
This is certainly not a coincidence, as the left-hand side of Eq.~\ref{eq:mig} is related to $\Delta\tilde{\mu}_{+,\ssc}$, which corresponds to the applied potential $\Delta V$, and that of Eq.~\ref{eq:diff} is proportional to the chemical
potential of the salt. In both cases, these are the quantities which can be measured in experiments. 

When \added{taking the ratio between} Eq.~\ref{eq:mig} and Eq.~\ref{eq:diff}, one \added{can relate the chemical potential change of the salt $\Delta \mu_\textrm{salt} = \Delta \mu_{+,\ssc} + \Delta \mu_{-,\ssc}$ to the corresponding potential difference $\Delta \phi_\mathrm{salt}$ defined as:
\begin{equation}
     \Delta \phi_\mathrm{salt} = \frac{RT}{zF}\ln\left(\frac{\gamma_{\pm, a}\,m_a}{\gamma_{\pm, c}\,m_c}\right)
\end{equation}
, where $\gamma_{\pm}$ is the mean activity coefficient and $m$ is the molal salt concentration. Then, this leads to the expression of $\Delta \phi_\mathrm{salt}$ in terms of the applied potential and the Onsager coefficients as}
\begin{equation}
    \Delta \phi_\mathrm{salt}=\frac{RT}{2zF}\ln\left(\frac{a_a}{a_c}\right) = \left(\frac{\Omega^{0}_{--}-\Omega^{0}_{+-}}{2\Omega^{0}_{--}}\right)\Delta V
\end{equation}
, where $a_a$ and $a_c$ are the salt activities near anode and cathode respectively. This recovers the dilute limit ($\Omega^{0}_{+-} \rightarrow  0$) in which $\Delta \phi_\mathrm{salt}$ equals to half of the applied potential at the steady-state~\cite{1987_BruceVincent}. 

Therefore, what we reveal here
is yet another example of the Gibbs-Guggenheim principle~\cite{doi:10.1021/j150300a003, Pethica:2007ki}, stating that chemical potentials of individual ions are a mathematical 
construct and cannot be measured experimentally without extrathermodynamic assumptions (e.g. the choice of $x$ in this case). Nevertheless, what matters is that $t_{+, \ssc}$ is a well-defined quantity and its derivation in terms of Onsager coefficients does not need to involve any of these extrathermodynamic assumptions. 

Now, we are ready to apply Eq.~\ref{tss_eq2} to the PEO-LiTFSI system by using the MD simulations. The simulations were performed
using GROMACS\cite{2015_AbrahamMurtolaEtAl} package and the General AMBER Force
Field (GAFF)\cite{2004_WangWolfEtAl} at 157$^\circ$C because of a much higher glass transition temperature $T_g$ found in the simulation system~\cite{10.1021/acs.jpcb.0c05108}. Further details regarding the simulation setup and
the force field parameterization can be found in the previous works~\cite{10.1021/acs.jpcb.0c05108, 10.1021/acs.jpclett.1c02474}.

Onsager coefficients under the barycentric RF (denoted as M) can be readily computed
 from the displacement
correlations as a function of time $t$:

\begin{equation}
  \Omega_{ij}^{\mathrm{M}}=%
  \lim _{t \to \infty} \frac{\beta}{6 VN_{\mathrm{A}}^2 t}%
  \left\langle
    \Delta \mathbf{r}_{i}^{\mathrm{M}}(t) \cdot \Delta \mathbf{r}_{j}^{\mathrm{M}}(t)
  \right\rangle
\end{equation}
, where $\beta$ is the inverse temperature, $V$ is the system volume,
$N_{\mathrm{A}}$ is the Avogadro number, and $\Delta\mathbf{r}_{i}^{\mathrm{M}}(t)$ is the total displacement of 
species $i$ over a time interval $t$. The long-time limit is estimated by
fitting the correlation as a linear function over the interval 10-20 ns, which
was shown to reach the diffusion regime in the previous work~\cite{2022_ShaoGudlaEtAla}.

To compute $t_{+, \ssc}$, it is necessary to convert Onsager
coefficients from the barycentric RF to the solvent-fixed RF (denoted as 0) using the
transformation rule~\cite{2022_ShaoGudlaEtAla}:
\begin{align}
  \Omega_{ij}^{0}=& \sum_{k, l \neq 0} A_{ik}^{\mathrm{0M}} \Omega_{k l}^{\mathrm{M}} A_{j l}^{\mathrm{0M}} \nonumber\\
  \begin{split}
                 =& \lim_{t\to\infty}\frac{\beta}{6VN_\mathrm{A}^2t} \cdot \\
                  & \left\langle \left(\sum_{k \neq 0} A_{ik}^{\mathrm{0M}} \Delta\mathbf{r}_{k}^\mathrm{M}(t)\right)
                   \cdot
                   \left(\sum_{l \neq 0} A_{jl}^{\mathrm{0M}} \Delta\mathbf{r}_{l}^\mathrm{M}(t) \right)
                   \right\rangle
  \end{split}\label{eq:corr}
\end{align}
, where $A_{ij}^{\mathrm{0M}}$ is the matrix converting the $n-1$ independent fluxes
in an $n$-component system from the barycentric RF to the
solvent-fixed one. Eq.~\ref{eq:corr} shows how this transformation relates to
the correlations of ions. The detailed derivation of such relation from constraints of the
fluxes and driving forces can be found elsewhere\cite{1984_GrootMazurEtAl,1966_Miller}. An important implication from these
works is that when the response
relation as shown here is applied in other RFs, it also entails a corresponding RF transformation of these driving forces.  

\begin{figure}[ht]
    \includegraphics[width=\linewidth]{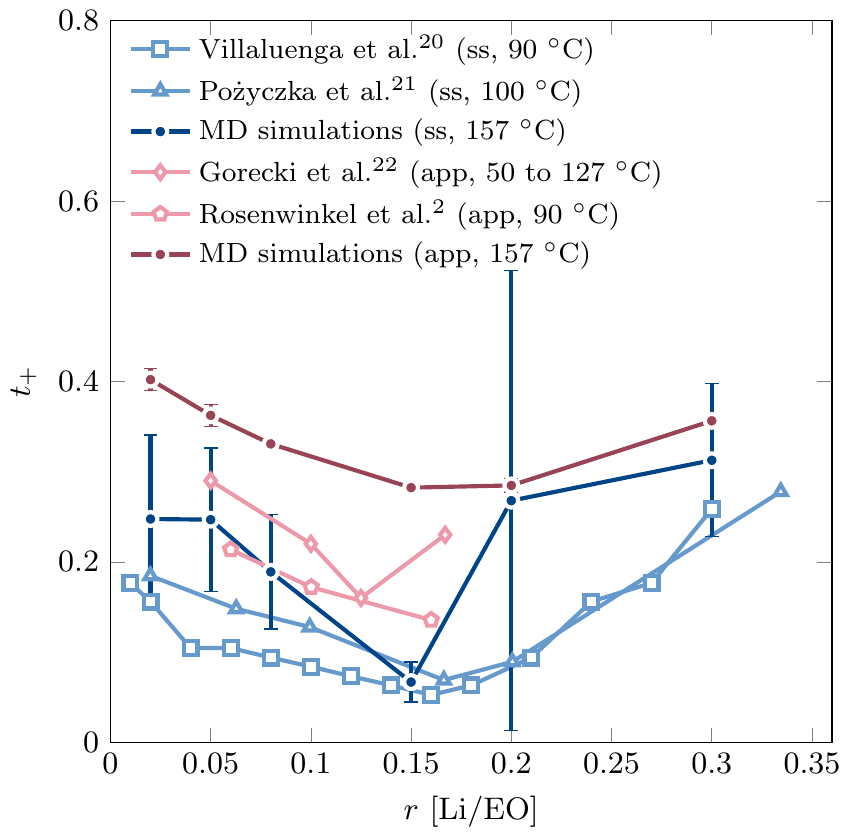}
    \caption{Experimental and simulation results of Bruce-Vincent
      transference numbers $t_{+,\ssc}$ and apparent transference numbers
      $t_{+,\app}$ for the PEO-LiTFSI system. The experimental data were taken from Ref.
      \citenum{2018_VillaluengaPeskoEtAl,%
        2017_PozyczkaMarzantowiczEtAl,%
        1995_GoreckiJeanninEtAl,%
        2019_RosenwinkelSchoenhoff}.}
\label{PEO_tss}
\end{figure}

The results of computed $t_{+, \ssc}$ for the PEO-LiTFSI system in comparison
with experiments are shown in Fig.~\ref{PEO_tss}, together with those of
$t_{+, \app}$. It is found that $t_{+,\ssc}$ is always positive as expected
since both diagonal Onsager coefficients and the determinant are positive; the
same is true for $t_{+,\mathrm{app}}$ and the two quantities approach each
other at the dilute condition. In addition, the relation
$t_{+,\mathrm{ss}}<t_{+,\mathrm{app}}$ seems to hold for the entire range of
concentration, in both simulation and experiment.

Given that the measurements of Bruce-Vincent transference numbers are accessible
in most labs and $t_{+, \ssc}$ does reflect a defined facet of the ion-ion
correlations, this work provides a way to calibrate the MD simulations of
polymer electrolytes and concentrated electrolytes alike by reproducing
$t_{+, \ssc}$. Through quantitative comparison and calibration, this will then
allow us to use MD simulations as a predictive tool to obtain the true
transference number $t^0_+$ for new types of polymer platforms beyond PEO.

\section*{acknowledgements}
  This work has been supported by the Swedish Research Council (VR), grant no.
  2019-05012. The authors thank funding from the Swedish National Strategic
  e-Science program eSSENCE, STandUP for Energy and BASE (Batteries Sweden). The
  simulations were performed on the resources provided by the National Academic
  Infrastructure for Supercomputing in Sweden (NAISS) at PDC.

\section*{Data Availability Statement}
The data that supports the findings of this study are available within the article.

%

\end{document}